\documentclass[%
 reprint,
nofootinbib,
 amsmath,amssymb,
 aps,
 prl,
]{revtex4-2}

\usepackage{graphicx}
\usepackage{dcolumn}
\usepackage{bm}
\usepackage{miller}
\usepackage{setspace}
\usepackage{siunitx}
\usepackage{nicefrac}
\usepackage{enumerate}
\usepackage{multirow}
\usepackage[]{todonotes}
\usepackage{tabularx}
\usepackage{makecell} 
\usepackage{tablefootnote}
\usepackage{amssymb}
\usepackage{amsmath}
\usepackage{comment}
\usepackage{hyperref}  
\usepackage{xr}  
\usepackage{soul}
\usepackage{float}
\floatstyle{plaintop}
\restylefloat{table}
\usepackage{enumitem}
\usepackage[english]{babel}
\usepackage{color}

\begin{document}

\preprint{prl}

\title{Intrinsic linewidths of confined phonons in few-layer hBN}

\author{Aleksandar Radi\'{c}}\email{ar2071@cam.ac.uk}
\affiliation{Cavendish Laboratory, Department of Physics, University of Cambridge, JJ Thomson Ave, Cambridge, United Kingdom}%

\author{Jack Kelsall}
\affiliation{Cavendish Laboratory, Department of Physics, University of Cambridge, JJ Thomson Ave, Cambridge, United Kingdom}%

\author{Akshay Rao}
\affiliation{Cavendish Laboratory, Department of Physics, University of Cambridge, JJ Thomson Ave, Cambridge, United Kingdom}%

\begin{abstract}
\noindent Understanding lattice vibrations in two-dimensional (2D) materials is essential for controlling thermal transport, mechanical response, and energy dissipation in nanoscale devices. However, the intrinsic lifetimes of low-energy phonon modes, particularly those that are optically silent, remain largely unexplored. Here we use helium-3 spin-echo spectroscopy to resolve low-energy phonons at the surface of hexagonal boron nitride (hBN) and measure their intrinsic linewidths. We observe the flexural and Rayleigh wave modes and extract the bending rigidity of a quasi-freestanding hBN monolayer. We further report the simultaneous observation of multiple surface-confined interlayer shear modes whose energies agree closely with linear-chain model predictions. By resolving their intrinsic linewidths, we demonstrate a strong confinement-induced reduction in phonon lifetimes, with a near order of magnitude increase in linewidth between the four- and two-layer modes. The temperature dependence of the linewidths indicates that phonon–phonon scattering dominates between $160-\SI{360}{\kelvin}$, while the systematic broadening with decreasing layer number reveals the impact of confinement on phonon decay. These results reveal how reduced dimensionality affects the decay of interlayer shearing modes in hBN, providing direct insight into the phonon lifetimes, confinement effects, and dissipation pathways that govern the dynamical behaviour of two-dimensional materials.
\end{abstract}

\maketitle
Understanding lattice vibrations in two-dimensional (2D) materials is central to controlling thermal transport, mechanical response, and energy dissipation in nanoscale devices. In layered crystals, reduced dimensionality and weak interlayer coupling give rise to vibrational modes with no direct bulk analogue, including flexural, interlayer shear, and layer-breathing modes. These low-energy excitations govern heat flow, mechanical stability, and phonon-mediated energy dissipation in atomically thin systems. However, despite their importance, the intrinsic lifetimes of these modes remain largely unexplored.

Hexagonal boron nitride (hBN) occupies a unique position among 2D materials. Its large bandgap ($\sim\SI{6}{\electronvolt}$), chemical stability, and atomically flat surface make it a popular dielectric and encapsulation material in 2D semiconductor devices. However, these same properties render its low-energy vibrational spectrum difficult to access experimentally. Optical probes such as Raman spectroscopy are intrinsically limited in hBN due to its non-resonant character, weak scattering cross-sections, and susceptibility to laser-induced heating, which is particularly severe in few-layer samples\cite{Reich2005ResonantNitride,Stenger2017LowCrystals}. As a result, both the frequencies and linewidths of several fundamental phonon modes in hBN, including flexural, layer-breathing, and lower-order interlayer shear modes, remain experimentally unresolved\cite{Michel2012TheoryCrystals,Tan2012TheGraphene}.

In the case of interlayer shear modes (ISMs), the linear-chain model predicts $N-1$ modes for an $N$-layer system\cite{Tan2012TheGraphene}, yet only the highest-frequency mode has been observed experimentally in a given sample thickness. The flexural and layer-breathing modes are optically silent at the centre of the Brillouin zone due to symmetry constraints\cite{Tan2012TheGraphene,Michel2012TheoryCrystals,Stenger2017LowCrystals}. To our knowledge, experimental linewidths of interlayer shear modes (ISMs) in few-layer hBN have not been measured previously. The absence of experimental characterisation leaves a significant gap in our understanding of vibrational dynamics and energy dissipation in layered materials.

Helium-3 spin-echo (HeSE) spectroscopy provides a complementary surface-sensitive probe of lattice dynamics that is uniquely suited to addressing the characterisation of surface-confined and few-layer phonons. As a charge-neutral, low-energy probe, helium scatters from the surface electron density and measures phonon dispersions and linewidths without injecting charge carriers or optical power into the lattice\cite{Alexandrowicz2007HeliumResolution,Jardine2004Ultrahigh-ResolutionLandscapes}. Crucially, the technique is sensitive to phonons irrespective of optical selection rules, enabling direct measurements of vibrational modes that are inaccessible to conventional spectroscopies\cite{Liu2024ExperimentalShortening,Liu2024DistinguishingMeasurements,Radic2025DefectmodifiedMoS2}. In addition, the ultrahigh energy resolution of HeSE enables direct access to intrinsic phonon linewidths, and therefore lifetimes, under equilibrium conditions.

In this Letter, we use HeSE to resolve the low-energy phonons at the surface of bulk hBN and directly measure their intrinsic lifetimes. We observe the flexural and Rayleigh wave (RW) modes and use the flexural branch to extract the mechanical properties of a quasi-freestanding hBN monolayer using simple harmonic membrane–substrate coupling theory\cite{Amorim2013FlexuralSubstrate,AlTaleb2015HeliumFoil,Radic2025DefectmodifiedMoS2}. We then report the simultaneous observation of three surface-confined ISMs corresponding to four-, three- and two-layer oscillations, whose energies agree closely with linear-chain model (LCM) predictions\cite{Tan2012TheGraphene}. By resolving their intrinsic linewidths, and evaluating the defect-phonon and phonon-phonon contributions, we find nearly an order of magnitude decrease from four- to two-layer modes. Together, these results highlight the importance of continued experimental research into phonon lifetimes in two-dimensional systems.

\subsection{A quasi-freestanding monolayer}


In 2015 Al Taleb \emph{et al.} introduced inelastic helium scattering (HAS) as a new measurement technique for the bending rigidity of 2D materials using the flexural phonon mode\cite{AlTaleb2015HeliumFoil}. The theoretical treatment by Amorim \& Guinea\cite{Amorim2013FlexuralSubstrate} and the surface sensitivity of a helium atom probe allows us to treat a 2D material with weak substrate coupling as a quasi-freestanding monolayer\cite{AlTaleb2015HeliumFoil,Buchner2018BendingSilica, Tmterud2022TemperatureGraphene,Radic2025DefectmodifiedMoS2}. In the following section, we demonstrate that the topmost layer of hBN also vibrates as a quasi-freestanding monolayer by measuring the flexural and Rayleigh wave modes.

\begin{figure}[h]
    \centering
    \includegraphics[width=76mm]{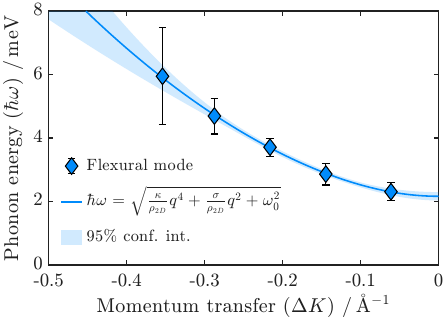}
    \caption{Dispersion of the flexural phonon mode at the surface of hBN along the $\overline{\Gamma\mathrm{M}}$ direction. The flexural mode arises from simple harmonic coupling between a weakly bound (quasi-freestanding) monolayer and a substrate, and is well-described by an approximately quadratic dispersion (Eqn.\ref{eqn:flex_dispersion})\cite{Amorim2013FlexuralSubstrate,Buchner2018BendingSilica}. Individual points and uncertainties were extracted from Lorentzian fitting to energy transfer spectra (Figure \ref{fig:flex_waterfall}) at sample temperature $T=\SI{360}{\kelvin}$ and tilt angle $\alpha=\SI{135}{\degree}$. All data shown here appear as annihilation $(+\Delta E)$ features in energy transfer spectra. The shaded region represents the 95\% confidence interval of the fit to Eqn. \eqref{eqn:flex_dispersion}.}
    \label{fig:dispersions}
\end{figure}

We first identify the low-energy surface phonon modes of hBN using HeSE. In Figure \ref{fig:dispersions}, we present the dispersions of the flexural and Rayleigh wave (RW) modes along the high-symmetry $\overline{\Gamma \mathrm{M}}$ direction.

The flexural mode (Figure \ref{fig:dispersions}) is well described by the dispersion expected for a two-dimensional membrane weakly coupled to a substrate as a harmonic oscillator\cite{Amorim2013FlexuralSubstrate},
\begin{equation}
\omega(q)^2=
\frac{\kappa}{\rho_{\mathrm{2D}}}q^4+
\frac{\sigma}{\rho_{\mathrm{2D}}}q^2+
\omega_0^2,
\label{eqn:flex_dispersion}
\end{equation}
where $\kappa$ is the bending rigidity, $\sigma$ the effective membrane tension, and $\omega_0$ accounts for substrate pinning. The extracted parameters ($\kappa = \SI{0.7(0.6)}{\electronvolt}$, $\sigma = \SI{2.8(0.8)}{\newton\per\metre}$, $g = \SI{8.2(1.0)e18}{\newton\per\metre\cubed}$) lie within the range reported for monolayer hBN and weakly coupled two-dimensional systems\cite{Amorim2013FlexuralSubstrate,Michel2011PhononNitride,Mariani2008FlexuralGraphene,Ong2011EffectGraphene}. One can also evaluate the Young's modulus $Y$ using $\kappa=\frac{Yh^3}{12(1-\nu^2)}$ with Poisson's ratio $\nu$ and layer thickness $h$, however this implies knowledge of $h$. Just as in the case of graphene, however, $h$ is ambiguous for hBN and there are a range of plausible values that give vastly different values for Young's modulus. This is known as the `Yakobson Paradox' and further discussion can be found in Refs. \cite{Buchner2018BendingSilica,Wang2005SizeNanotubes,Pine2014VibrationalStrain, Huang2006ThicknessNanotubes}.

We also observe the Rayleigh wave (RW) mode, although only under a distinct scattering condition. In HeSE, the tilt angle $\alpha$ sets the projection of the measurement through the wavelength-intensity matrix (WIM) and therefore influences which spectral features are most clearly resolved\cite{Alexandrowicz2007HeliumResolution}.  In contrast to the flexural mode, which was measured at $\alpha=\SI{135}{\degree}$, $T=\SI{360}{\kelvin}$, the RW mode was resolved at $\alpha=\SI{125}{\degree}$, $T=\SI{200}{\kelvin}$. Energy transfer spectra are shown in Figure \ref{fig:rw_waterfall}. Mode selectivity arises from the scan-curve geometry of the HeSE measurement: changing the tilt angle $\alpha$ changes the projection through the WIM and therefore the portion of the surface dynamical response that is sampled. The tilt angle as a measurement parameter is specific to HeSE because it is set by the ratio of the incoming and outgoing solenoid currents $\tan\alpha=\frac{I_i}{I_f}$, and therefore has no direct analogue in conventional HAS instruments. As a result, a mode that is only clearly resolved under a particular WIM projection may be effectively hidden due to severe broadening in conventional energy-loss measurements\cite{Alexandrowicz2007HeliumResolution}. The RW branch is therefore visible at the $\SI{125}{\degree}$ tilt angle, while under the $\SI{135}{\degree}$ condition its contribution is not resolved above the quasi-elastic feature, multi-phonon background, and the flexural mode. In Figure \ref{fig:wavelength_intensity} we use a phenomenological wavelength-intensity representation, constrained by experimental data, to illustrate how the tilt angle changes the projected spectral weight in the measured spectra. The extracted RW dispersion (Figure \ref{fig:rw_dispersion}) is approximately linear over the measured range with group velocity $v_{\mathrm{RW}}=\SI{2.30(0.38)}{\kilo\metre\per\second}$.

The flexural dispersion provides direct evidence that the hBN surface behaves as a weakly supported two-dimensional membrane. Its form and fitted parameters are consistent with a quasi-freestanding monolayer coupled harmonically to the underlying crystal, rather than with a bulk acoustic mode. Together with the separately resolved RW branch, this shows that HeSE probes the surface-localised phonons of hBN. We therefore assign the measured modes to the topmost hBN layer, consistent with previous helium atom scattering studies of layered materials\cite{AlTaleb2015HeliumFoil,Buchner2018BendingSilica,Tmterud2025HeliumMaterials,Radic2025DefectmodifiedMoS2}. This motivates the interpretation of the confined interlayer shear modes and their linewidths/lifetimes discussed below.

\begin{figure*}
    \centering
    \includegraphics[width=163mm]{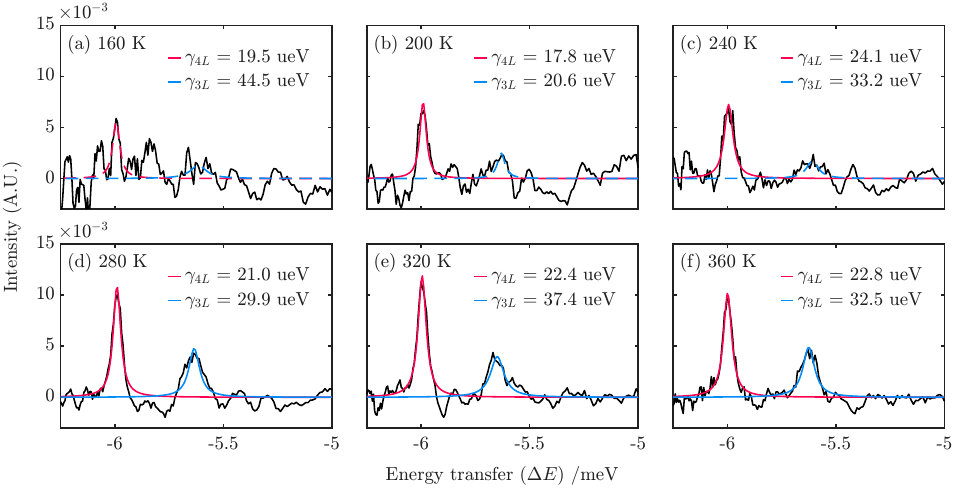}
    \caption{Enegy transfer spectra of the 3L (blue) and 4L (red) interlayer shearing modes (ISMs) with single Lorentzian fits for sample temperatures $160-\SI{360}{\kelvin}$ and tilt angle $\alpha=\SI{129.1}{\degree}$. Linewidth for each fit is shown in the legend of the respective panel. In Figure \ref{fig:ISM_linewidths_temp} we analyse the temperature dependence of these linewidths alongside the fitting of the 2L mode. Fits with high uncertainty are shown with dashed lines and are not used in the subsequent analysis. A 2\textsuperscript{nd}-order polynomial background subtraction was applied to account for multi-phonon background contributions to intensity followed by a Savitsky-Golay filter (1\textsuperscript{st}-order with window width 7) is used to remove high-frequency noise.}
    \label{fig:ISM_linewidths}
\end{figure*}

\subsection{Confined interlayer shear modes}

In layered materials, shearing between adjacent planes gives rise to interlayer shearing modes (ISMs), whose energies and symmetries are strongly dependent on the number of coupled layers\cite{Michel2012TheoryCrystals}. At a surface or interface, the finite number of mechanically coupled layers leads to discrete, surface-confined ISMs. For a sample with $N$ layers, the linear-chain model (LCM) predicts $N-1$ modes\cite{Tan2012TheGraphene}, however these lower-indexed modes have remained elusive to experimental measurement because their scattering cross-sections decrease by orders of magnitude due to symmetry and stacking-dependent changes in polarizability\cite{Luo2015StackingModel}. In addition, mutual exclusion of Raman and IR accessibility further suppresses observable modes depending on layer parity.

In Figure \ref{fig:ISM_linewidths}, we resolve the 4L, 3L and 2L modes simultaneously, corresponding to shear oscillations of four-, three- and two-layer systems, respectively. The measured energies agree with LCM predictions (Figure \ref{fig:confinement}, End Matter) to within $1\%$. We find relative intensities of $4L\colon3L = 2.5\colon1$ and $3L\colon2L = 5.1\colon1$ ($4L\colon2L = 12\colon1$), consistent with the expected suppression of lower-index modes. Notably, no feature is observed at the energy expected for the 5L shear mode, suggesting that the intrinsic vertical coherence of these modes is limited to approximately four layers.

The measured ISM energies are in excellent agreement with the LCM for the observed modes. In contrast, Raman measurements of few-layer hBN systematically report higher frequencies than predicted for samples thinner than approximately five layers\cite{Stenger2017LowCrystals}. The inset of Fig. \ref{fig:confinement} (End Matter) compares Raman and HeSE (this work) measurements to the LCM, and shows that while Raman frequencies deviate upward from the model, all HeSE values lie within $1\%$ of LCM predictions. The difference is likely caused by probe-induced heating in optical measurements, which is particularly severe in few-layer hBN due to weak optical absorption and poor heat dissipation\cite{Schue2016CharacterizationLayers,Stenger2017LowCrystals}. As a charge-neutral, low-energy probe, HeSE avoids such effects and provides direct access to the intrinsic vibrational spectrum of these optically silent modes.

\subsection{Linewidths of interlayer shear modes}

Optical measurements of interlayer shearing modes (ISMs) primarily focus on frequency shifts, while linewidths are typically instrument response-limited or obscured by probe-induced effects\cite{Stenger2017LowCrystals}. In particular, low-frequency ISMs exhibit weak Raman scattering cross-sections and lie close to the Rayleigh filter edge, making reliable extraction of intrinsic linewidths challenging. Furthermore, laser-induced heating and inhomogeneous broadening can dominate the observed linewidths\cite{Schue2016CharacterizationLayers,Stenger2017LowCrystals}. As a result, equilibrium lifetimes of ISMs remain elusive.

\begin{figure}
    \centering
    \includegraphics[width=78mm]{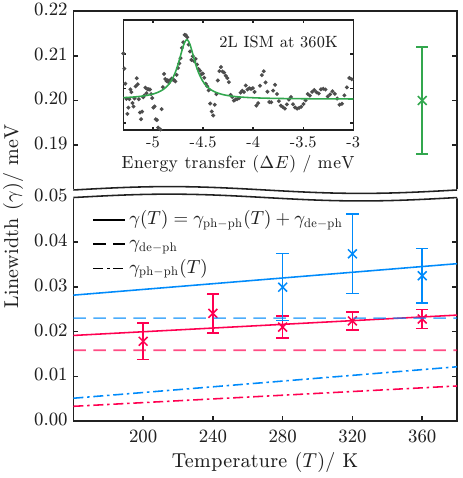}
    \caption{Linewidths of the 2--4L interlayer shearing modes (ISM) at the surface of bulk hBN from $160-\SI{360}{\kelvin}$. Inset shows the energy transfer spectrum of the 2L mode fit with a Lorentzian at $\SI{360}{\kelvin}$ and tilt angle $\alpha=\SI{129.1}{\degree}$ yielding frequency $\omega_{2L}=\SI{4.65}{\milli\electronvolt}$ and linewidth $\gamma_{2L}=\SI{210(40)}{\micro\electronvolt}$. The linear relationship between linewidth and temperature for the 3L and 4L modes indicates dominant phonon-phonon scattering while the constant offset corresponds to defect-phonon scattering\cite{Liu2024UltrahighPhononics}. The phonon-phonon ($\gamma_{\mathrm{ph-ph}}$) and defect-phonon ($\gamma_{\mathrm{de-ph}}$) linewidth contributions are shown as dot-dashed and dashed lines, respectively, for the 3L and 4L modes. We neglect the electron-phonon contribution due to the insulating nature of hBN.}
    \label{fig:ISM_linewidths_temp}
\end{figure}

In Figure \ref{fig:ISM_linewidths_temp}, we report the linewidths of the 4L, 3L and 2L ISMs. At $\SI{360}{\kelvin}$, we obtain $\gamma_{4L}=\SI{22.8(2.1)}{\micro\electronvolt}$, $\gamma_{3L}=\SI{32.5(6.1)}{\micro\electronvolt}$ and $\gamma_{2L}=\SI{210(40)}{\micro\electronvolt}$, corresponding to lifetimes of $\tau_{4L}=\SI{28.9(2.7)}{\pico\second}$, $\tau_{3L}=\SI{20.3(3.8)}{\pico\second}$ and $\tau_{2L}=\SI{3.3(0.6)}{\pico\second}$. For the three- and four-layer modes, their linewidths increase linearly with temperature, $\gamma_{\mathrm{ph-ph,3L}}(T)=\SI{32}{\nano\electronvolt\per\kelvin}$ and $\gamma_{\mathrm{ph-ph,4L}}(T)=\SI{21}{\nano\electronvolt\per\kelvin}$ respectively, indicating that phonon--phonon scattering dominates in the $160-\SI{360}{\kelvin}$ range\cite{Liu2024UltrahighPhononics}. A finite intercept at zero temperature suggests an additional temperature-independent contribution that we attribute to defect--phonon scattering, $\gamma_{\mathrm{de-ph, 3L}}=\SI{23.0}{\micro\electronvolt}$ and $\gamma_{\mathrm{de-ph, 4L}}=\SI{15.8}{\micro\electronvolt}$, while electron--phonon scattering is neglected due to the insulating nature of hBN. It is interesting to note that $\gamma_{\mathrm{de-ph}}>\gamma_{\mathrm{ph-ph}}$ for temperatures at least as high as $\SI{360}{\kelvin}$, suggesting that defect-phonon scattering remains the dominant lifetime-limiting mechanism  across a range that extends well above typical device operating temperatures, even in hBN - a material known for its comparatively low defect density among 2D crystals.

The linewidths increase systematically with decreasing layer number. The modest increase in linewidth from 4L to 3L is quantitatively consistent with confinement-enhanced anharmonicity: reducing the number of participating layers increases the vibrational amplitude per layer as $u\propto N^{-1/2}$, leading to $\gamma_{\mathrm{ph-ph}}\propto u^2\propto N^{-1}$. The ratios of the phonon-phonon and defect-phonon contributions, $\gamma_{\mathrm{ph-ph,3L}}/\gamma_{\mathrm{ph-ph,4L}}\approx1.5$ and $\gamma_{\mathrm{de-ph,3L}}/\gamma_{\mathrm{de-ph,4L}}\approx1.5$, are in reasonable agreement with the $4/3$ scaling expected when considering only confinement-enhanced increases in oscillation amplitude. This indicates that the increase in linewidth from 4L to 3L can be described by confinement-enhanced anharmonicity.

In contrast, the substantially larger linewidth of the 2L mode cannot be explained by the previous scaling, increasing by nearly an order of magnitude compared to the 4L mode. The significant broadening of the bilayer mode indicates that it lies in a strongly confined regime where additional decay channels become available. Reduced symmetry, enhanced surface localisation, and increased sensitivity to disorder are expected to increase both phonon--phonon and defect--phonon scattering rates, leading to a pronounced reduction in phonon lifetimes.

The ability to resolve such narrow linewidths highlights the utility of HeSE for probing intrinsic phonon lifetimes. More broadly, these results demonstrate that few-layer confinement can strongly modify phonon decay, providing direct experimental access to dissipation mechanisms in two-dimensional materials.


\subsection{Conclusion}

We have used helium-3 spin-echo spectroscopy to measure the intrinsic low-energy vibrational spectrum at the surface of hexagonal boron nitride, accessing numerous phonon modes that have so far been inaccessible using standard techniques.

Characterisation of the flexural and Rayleigh wave dispersions confirm that the topmost hBN layer behaves as a quasi-freestanding monolayer harmonically coupled to the underlying bulk. Applying established continuum membrane–substrate theory to the flexural dispersion yields a bending rigidity, effective tension, and substrate pinning frequency consistent with monolayer hBN and with weakly coupled two-dimensional systems more broadly, demonstrating that bulk crystals can serve as a platform for extracting single-layer mechanical properties.

We simultaneously observe the 2L, 3L, and 4L interlayer shearing modes, whose energies lie within 1\% of linear-chain model predictions. This agreement stands in contrast to previous measurements which systematically overestimate the modes' frequencies - an effect we attribute to probe-induced heating. As a charge-neutral, $\sim\SI{8}{\milli\electronvolt}$ probe, HeSE avoids probe heating entirely and accesses equilibrium vibrational behaviour.

For the 3L and 4L modes, we decompose their linewidths into defect-phonon and phonon-phonon contributions, neglecting the electron-phonon contribution due to the insulating nature of hBN. Their linewidths increase linearly with temperature, identifying phonon-phonon scattering as the source of the temperature-dependent broadening across the $160$-$\SI{360}{\kelvin}$ range. However, the temperature-independent defect-phonon contribution remains larger throughout the measured range, with the fitted phonon-phonon contribution predicted to overtake it past $\SI{700}{\kelvin}$.

Both the phonon-phonon and defect-phonon contributions increase from the 4L to the 3L mode in a manner consistent with confinement scaling. Their respective ratios, $\gamma_{\mathrm{ph-ph,3L}}/\gamma_{\mathrm{ph-ph,4L}}\approx1.5$ and $\gamma_{\mathrm{de-ph,3L}}/\gamma_{\mathrm{de-ph,4L}}\approx1.5$, are close to the $4/3$ value expected from $\gamma\propto N^{-1}$. In contrast, the 2L linewidth at $\SI{360}{\kelvin}$ is nearly an order of magnitude larger than that of the 4L mode, indicating a strongly confined regime with enhanced decay.

These measurements provide direct access to equilibrium phonon linewidths and decay mechanisms that are difficult to resolve otherwise. The ability to resolve these intrinsic phonon properties opens new opportunities to study how confinement and disorder affect mechanical response and energy dissipation in two-dimensional materials and devices.

\subsection{Data Availability}
Supporting data and code will be made available upon publication.

\subsection{Acknowledgments}
This work was supported by a UK Research and Innovation (UKRI)
Frontier Research Grant (EP/Y015584/1). The work was performed in part at CORDE, the Collaborative R\&D Environment established to provide access to physics related facilities at the Cavendish Laboratory, University of Cambridge and EPSRC award EP/T00634X/1. The authors thank Boyao Liu, Sam Lambrick, Bill Allison, and John Ellis for useful discussions.

\clearpage
\appendix

\renewcommand{\thefigure}{A\arabic{figure}}
\setcounter{figure}{0}

\renewcommand{\thetable}{A\Roman{table}}
\setcounter{table}{0}

\renewcommand{\theequation}{A\arabic{equation}}
\setcounter{equation}{0}

\section{Energy transfer spectra for flexural and Rayleigh wave modes}
\begin{figure}[H]
\centering
\includegraphics[width=67mm]{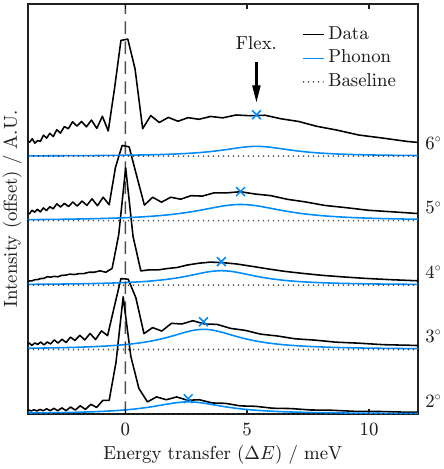}
\caption{Energy transfer spectra showing the flexural mode in quasi-freestanding hBN at sample temperature $T=\SI{360}{\kelvin}$ with a tilt angle $\alpha=\SI{135}{\degree}$ along the $\overline{\Gamma\mathrm{M}}$ direction. The dispersion relation of this mode is shown in Figure \ref{fig:dispersions}. Linewidths of the flexural mode could not be reliably extracted due to overlap with the quasi-elastic peak and multi-phonon background, and are therefore not analysed further. Fitted model includes a convex 2\textsuperscript{nd}-order polynomial to capture multi-phonon background, and a Lorentzian for the each of the quasi-elastic ($\Delta E=0$) and inelastic (phonon) features. Only the phonon component of the model is shown for visual clarity.}
\label{fig:flex_waterfall}
\end{figure}

\begin{figure}
    \centering
    \includegraphics[width=67mm]{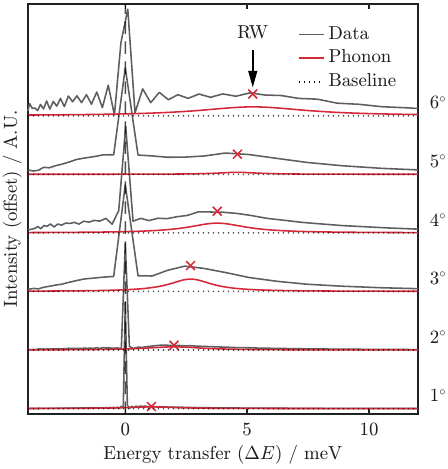}
    \caption{Energy transfer spectra showing the Rayleigh wave mode in quasi-freestanding hBN at sample temperature $T=\SI{200}{\kelvin}$ with a tilt angle $\alpha=\SI{125}{\degree}$ along the $\overline{\Gamma\mathrm{M}}$ direction. We note that the Rayleigh wave was only observed at this tilt angle, whereas all other measurements in the current work use $\alpha=\SI{135}{\degree}$. We discuss how selection of tilt angle in HeSE affects our ability to resolve different phonon modes in Figure \ref{fig:wavelength_intensity} using the wavelength-intensity formalism. Linewidths of the rayleigh wave could not be reliably extracted due to overlap with the quasi-elastic peak and multi-phonon background, and are therefore not analysed further. Fitted model includes a convex 2\textsuperscript{nd}-order polynomial to capture multi-phonon background, and a Lorentzian for the each of the quasi-elastic ($\Delta E=0$) and inelastic (phonon) features. Only the phonon component of the model is shown for visual clarity. Dispersion relation for the RW is shown in Figure \ref{fig:rw_dispersion}.}
    \label{fig:rw_waterfall}
\end{figure}

\begin{figure}
    \centering
    \includegraphics[width=67mm]{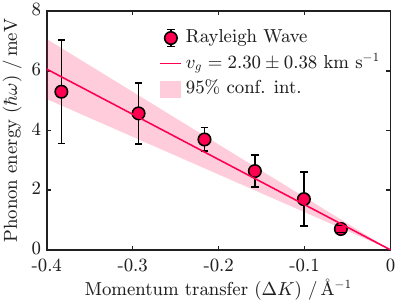}
    \caption{Dispersion of the Rayleigh wave (RW) phonon mode measured at the surface of hBN using HeSE with sample temperature $T=\SI{200}{\kelvin}$ along the $\overline{\Gamma\mathrm{M}}$ direction. We assume the RW to have approximately linear dispersion constrained through the origin yielding a group velocity $v_{g}=\SI{2.30(0.38)}{\kilo\metre\per\second}$. Energy transfer spectra for the RW are shown in Figure \ref{fig:rw_waterfall} where all phonon features appear as annihilation $(+\Delta E)$ peaks.}
    \label{fig:rw_dispersion}
\end{figure}

\begin{figure*}
    \centering
    \includegraphics[width=160mm]{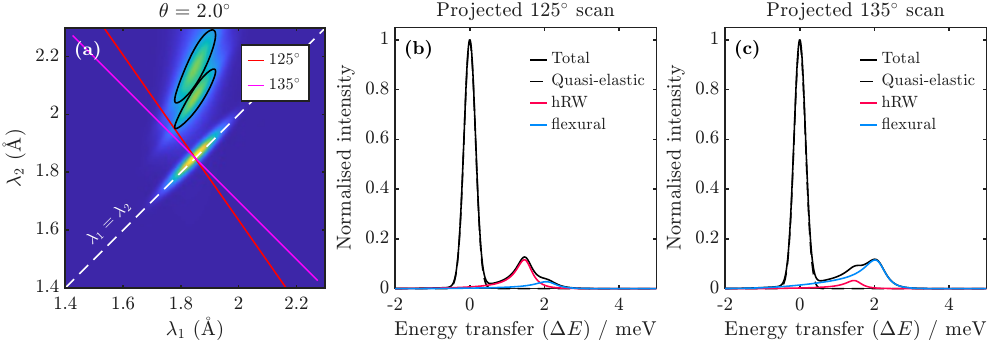}
    \caption{
    Model wavelength-intensity representation of the HeSE scattering geometry used to interpret the visibility of the surface acoustic modes. The intensity map (Panel (a)) is constructed from the experimentally determined flexural and Rayleigh-wave dispersions, using the measured incident beam energy ($\SI{8.0(0.5)}{\milli\electronvolt}$) and the corresponding sample-tilt-dependent momentum transfer. The quasi-elastic contribution is represented by a central ridge along $\lambda_1=\lambda_2$, while the phonons are included as finite-width spectral features satisfying the fitted dispersion relations, indicated schematically by the black ellipses. The relative amplitudes used in the projected spectra are phenomenological visibility factors: they illustrate that different wavelength-space scan directions project different fractions of each two-dimensional phonon feature into the measured one-dimensional spectrum, and do not represent intrinsic phonon spectral weights. Panel (a) shows normalised intensity on an arbitrary colour scale. Cuts through the matrix at $\alpha=\SI{125}{\degree}$ (red line) and $\alpha=\SI{135}{\degree}$ (magenta line) show how changing the scan direction changes the projection of the same two-dimensional response into the measured energy-transfer spectra (Panels (b), (c)). This provides a geometrical explanation for why the RW mode is most readily resolved under the $\SI{125}{\degree}$ condition, whereas the flexural mode is preferentially resolved under the $\SI{135}{\degree}$ condition. The model is a kinematic wavelength-intensity projection constrained by the measured dispersions and incident beam energy; it does not calculate absolute scattering intensities or mode-dependent matrix elements, and is therefore used only to illustrate the projection geometry that controls mode visibility. We note that the multi-phonon background, which becomes increasingly significant at larger values of momentum transfer, has not been modelled here.}
    \label{fig:wavelength_intensity}
\end{figure*}

\section{Comparison to Raman and Linear Chain Model (LCM)}
\begin{figure}[H]
    \centering
    \includegraphics[width=78mm]{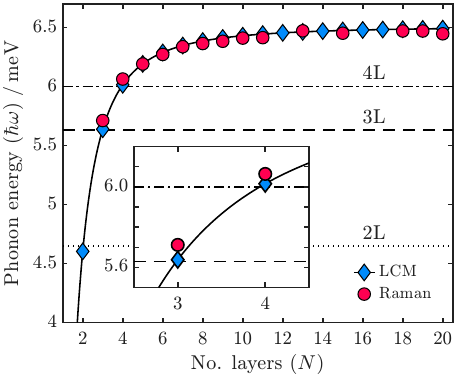}
    \caption{Linear-chain model (LCM) compared to measurements of the interlayer shearing (ISM) mode using Raman (circles) and HeSE (dashed lines). Raman measurements on exfoliated few-layer hBN flakes were conducted at $\SI{300}{\kelvin}$ by Stenger et al.\cite{Stenger2017LowCrystals} and adapted from Figure 3(c). HeSE phonon energies were extracted from energy transfer spectra in Figure \ref{fig:ISM_linewidths}.}
    \label{fig:confinement}
\end{figure}

Figure \ref{fig:confinement} compares the measured ISM energies to the linear-chain model (LCM)\cite{Tan2012TheGraphene} for interlayer shear vibrations, alongside Raman measurements of exfoliated few-layer hBN\cite{Stenger2017LowCrystals}. The LCM predicts a discrete set of shear modes whose energies depend sensitively on the number of layers and converge toward the bulk limit ($\hbar\omega_\mathrm{bulk}\sim\SI{6.5}{\milli\electronvolt}$). 

It is possible that laser-induced heating is indirectly responsible for the over-estimation of few-layer ISM frequencies, with Raman exceeding the HeSE/LCM values by $7\%$ and $3\%$ for 3L and 4L modes, respectively, while the 2L has not been recorded experimentally. Although heating is expected to reduce phonon frequency linearly in the high-temperature regime\cite{Liu2024UltrahighPhononics}, the heating can locally produce both inter- and intra-layer strain that ISMs are particularly sensitive to. Over-estimation of ISM frequencies suggests that second-order heating effects must be considered in 2D materials with poor optical cross-sections and/or thermal conductivity, like hBN.

\clearpage
\bibliography{references}

@article{Stenger2017LowCrystals,
    title = {{Low frequency Raman spectroscopy of few-atomic-layer thick hBN crystals}},
    year = {2017},
    journal = {2D Materials},
    author = {Stenger, I. and Schu, L. and Boukhicha, M. and Berini, B. and Pla{\c{c}}ais, B. and Loiseau, A. and Barjon, J.},
    number = {3},
    month = {6},
    pages = {031003},
    volume = {4},
    publisher = {IOP Publishing},
    url = {https://iopscience.iop.org/article/10.1088/2053-1583/aa77d4 https://iopscience.iop.org/article/10.1088/2053-1583/aa77d4/meta},
    doi = {10.1088/2053-1583/AA77D4},
    issn = {2053-1583}
}

@article{Amorim2013FlexuralSubstrate,
    title = {{Flexural mode of graphene on a substrate}},
    year = {2013},
    journal = {Physical Review B - Condensed Matter and Materials Physics},
    author = {Amorim, Bruno and Guinea, Francisco},
    number = {11},
    month = {9},
    pages = {115418},
    volume = {88},
    publisher = {American Physical Society},
    url = {https://journals.aps.org/prb/abstract/10.1103/PhysRevB.88.115418},
    doi = {10.1103/PHYSREVB.88.115418/DELIVERABLE/8B752DFF-558D-4C0E-841E-DED7641D1108},
    issn = {10980121},
    arxivId = {1304.6567}
}

@article{Michel2011PhononNitride,
    title = {{Phonon dispersions and piezoelectricity in bulk and multilayers of hexagonal boron nitride}},
    year = {2011},
    journal = {Physical Review B - Condensed Matter and Materials Physics},
    author = {Michel, K. H. and Verberck, B.},
    number = {11},
    month = {3},
    pages = {115328},
    volume = {83},
    publisher = {American Physical Society},
    url = {https://journals.aps.org/prb/abstract/10.1103/PhysRevB.83.115328},
    doi = {10.1103/PHYSREVB.83.115328/DELIVERABLE/41AF7DF3-F6D8-4628-8F8D-60F394DF9A1A},
    issn = {10980121}
}

@article{Mariani2008FlexuralGraphene,
    title = {{Flexural phonons in free-standing graphene}},
    year = {2008},
    journal = {Physical Review Letters},
    author = {Mariani, Eros and Von Oppen, Felix},
    number = {7},
    month = {2},
    pages = {076801},
    volume = {100},
    publisher = {American Physical Society},
    url = {https://journals.aps.org/prl/abstract/10.1103/PhysRevLett.100.076801},
    doi = {10.1103/PHYSREVLETT.100.076801/DELIVERABLE/A8DCCD47-F0FE-407E-A869-D5FB8DA55A13},
    issn = {00319007},
    arxivId = {0707.4350}
}

@article{Ong2011EffectGraphene,
    title = {{Effect of substrate modes on thermal transport in supported graphene}},
    year = {2011},
    journal = {Physical Review B - Condensed Matter and Materials Physics},
    author = {Ong, Zhun Yong and Pop, Eric},
    number = {7},
    month = {8},
    pages = {075471},
    volume = {84},
    publisher = {American Physical Society},
    url = {https://journals.aps.org/prb/abstract/10.1103/PhysRevB.84.075471},
    doi = {10.1103/PHYSREVB.84.075471/DELIVERABLE/76DE3D0F-B04A-49D0-98FA-817F899996C3},
    issn = {10980121}
}

@article{AlTaleb2015HeliumFoil,
    title = {{Helium diffraction and acoustic phonons of graphene grown on copper foil}},
    year = {2015},
    journal = {Carbon},
    author = {Al Taleb, Amjad and Yu, Hak Ki and Anemone, Gloria and Far{\'{i}}as, Daniel and Wodtke, Alec M.},
    month = {12},
    pages = {731--737},
    volume = {95},
    publisher = {Pergamon},
    doi = {10.1016/J.CARBON.2015.08.110},
    issn = {0008-6223}
}

@article{Liu2024ExperimentalShortening,
    title = {{Experimental Characterization of Defect-Induced Phonon Lifetime Shortening}},
    year = {2024},
    journal = {Physical Review Letters},
    author = {Liu, Boyao and Kelsall, Jack and Ward, David J and Jardine, Andrew P},
    number = {5},
    month = {1},
    pages = {56202},
    volume = {132},
    url = {https://link.aps.org/doi/10.1103/PhysRevLett.132.056202},
    doi = {10.1103/PhysRevLett.132.056202}
}

@phdthesis{Liu2024UltrahighPhononics,
    title = {{Ultrahigh resolution surface phononics}},
    year = {2024},
    author = {Liu, Boyao},
    publisher = {Apollo - University of Cambridge Repository},
    url = {https://www.repository.cam.ac.uk/handle/1810/374124},
    doi = {10.17863/CAM.112305}
}

@article{Michel2012TheoryCrystals,
    title = {{Theory of rigid-plane phonon modes in layered crystals}},
    year = {2012},
    journal = {Physical Review B - Condensed Matter and Materials Physics},
    author = {Michel, K. H. and Verberck, B.},
    number = {9},
    month = {3},
    pages = {094303},
    volume = {85},
    publisher = {American Physical Society},
    url = {https://journals.aps.org/prb/abstract/10.1103/PhysRevB.85.094303},
    doi = {10.1103/PHYSREVB.85.094303/DELIVERABLE/A3A73F83-463E-459D-BACE-A5DE9F751C76},
    issn = {10980121},
    arxivId = {1112.5544}
}

@article{Tan2012TheGraphene,
    title = {{The shear mode of multilayer graphene}},
    year = {2012},
    journal = {Nature Materials},
    author = {Tan, P H and Han, W P and Zhao, W J and Wu, Z H and Chang, K and Wang, H and Wang, Y F and Bonini, N and Marzari, N and Pugno, N and Savini, G and Lombardo, A and Ferrari, A C},
    number = {4},
    month = {2},
    pages = {294--300},
    volume = {11},
    publisher = {Springer Science and Business Media LLC},
    url = {http://dx.doi.org/10.1038/nmat3245},
    doi = {10.1038/nmat3245},
    issn = {1476-4660}
}

@article{Schue2016CharacterizationLayers,
    title = {{Characterization methods dedicated to nanometer-thick hBN layers}},
    year = {2016},
    journal = {2D Materials},
    author = {Schu{\'{e}}, Léonard and Stenger, Ingrid and Fossard, Frédéric and Loiseau, Annick and Barjon, Julien},
    number = {1},
    month = {12},
    pages = {015028},
    volume = {4},
    publisher = {IOP Publishing},
    url = {https://iopscience.iop.org/article/10.1088/2053-1583/4/1/015028 https://iopscience.iop.org/article/10.1088/2053-1583/4/1/015028/meta},
    doi = {10.1088/2053-1583/4/1/015028},
    issn = {2053-1583},
    arxivId = {1610.06858}
}

@article{Reich2005ResonantNitride,
    title = {{Resonant Raman scattering in cubic and hexagonal boron nitride}},
    year = {2005},
    journal = {Physical Review B - Condensed Matter and Materials Physics},
    author = {Reich, S. and Ferrari, A. C. and Arenal, R. and Loiseau, A. and Bello, I. and Robertson, J.},
    number = {20},
    month = {5},
    pages = {205201},
    volume = {71},
    publisher = {American Physical Society},
    url = {https://journals.aps.org/prb/abstract/10.1103/PhysRevB.71.205201},
    doi = {10.1103/PHYSREVB.71.205201/DELIVERABLE/6B7330BD-82AD-4AC9-82F0-AC91F2EE685C},
    issn = {10980121}
}

@article{Liu2024DistinguishingMeasurements,
    title = {{Distinguishing Quasiparticle-Phonon Interactions by Ultrahigh-Resolution Lifetime Measurements}},
    year = {2024},
    journal = {Physical Review Letters},
    author = {Liu, Boyao and Allison, William and Peng, Bo and Avidor, Nadav and Monserrat, Bartomeu and Jardine, Andrew P},
    number = {17},
    month = {4},
    pages = {176202},
    volume = {132},
    url = {https://link.aps.org/doi/10.1103/PhysRevLett.132.176202},
    doi = {10.1103/PhysRevLett.132.176202}
}

@article{Radic2025DefectmodifiedMoS2,
    title = {{Defect-modified acoustic phonons in a single layer of MoS2}},
    year = {2025},
    author = {Radic, Aleksandar and Liu, Boyao and Rao, Akshay and Lambrick, Sam M},
    month = {3},
    url = {https://arxiv.org/abs/2503.14464v3},
    arxivId = {2503.14464},
    journal = {arXiv}
}

@article{Luo2015StackingModel,
    title = {{Stacking sequence determines Raman intensities of observed interlayer shear modes in 2D layered materials-A general bond polarizability model}},
    year = {2015},
    journal = {Scientific Reports},
    author = {Luo, Xin and Lu, Xin and Cong, Chunxiao and Yu, Ting and Xiong, Qihua and Ying Quek, Su},
    month = {10},
    volume = {5},
    publisher = {Nature Publishing Group},
    doi = {10.1038/srep14565},
    issn = {20452322}
}

@article{Alexandrowicz2007HeliumResolution,
    title = {{Helium spin-echo spectroscopy: studying surface dynamics with ultra-high-energy resolution}},
    year = {2007},
    journal = {Journal of Physics: Condensed Matter},
    author = {Alexandrowicz, G and Jardine, A P},
    number = {30},
    month = {7},
    pages = {305001},
    volume = {19},
    url = {https://dx.doi.org/10.1088/0953-8984/19/30/305001},
    doi = {10.1088/0953-8984/19/30/305001},
    issn = {0953-8984},
    language = {english}
}

@article{Jardine2004Ultrahigh-ResolutionLandscapes,
    title = {{Ultrahigh-Resolution Spin-Echo Measurement of Surface Potential Energy Landscapes}},
    year = {2004},
    journal = {Science},
    author = {Jardine, Andrew P and Dworski, Shechar and Fouquet, Peter and Alexandrowicz, Gil and Riley, David J and Lee, Gabriel Y H and Ellis, John and Allison, William},
    number = {5678},
    month = {6},
    pages = {1790--1793},
    volume = {304},
    publisher = {American Association for the Advancement of Science (AAAS)},
    url = {http://dx.doi.org/10.1126/science.1098490},
    doi = {10.1126/science.1098490},
    issn = {1095-9203}
}

@article{Buchner2018BendingSilica,
    title = {{Bending Rigidity of 2D Silica}},
    year = {2018},
    journal = {Physical Review Letters},
    author = {B{\"{u}}chner, C. and Eder, S. D. and Nesse, T. and Kuhness, D. and Schlexer, P. and Pacchioni, G. and Manson, J. R. and Heyde, M. and Holst, B. and Freund, H. J.},
    number = {22},
    month = {5},
    volume = {120},
    publisher = {American Physical Society},
    doi = {10.1103/PHYSREVLETT.120.226101},
    issn = {10797114},
    pmid = {29906168}
}

@article{Tmterud2025HeliumMaterials,
    title = {{Helium atom scattering investigation of the boson peak presence in amorphous and crystalline two-dimensional materials}},
    year = {2025},
    journal = {Physical Review B},
    author = {T{\o}mterud, Martin and Eder, Sabrina D. and Hellner, Simen K. and B{\"{u}}chner, Christin and Heyde, Markus and Freund, Hans Joachim and Forti, Stiven and Convertino, Domenica and Coletti, Camilla and Manson, Joseph R. and Holst, Bodil},
    number = {20},
    month = {4},
    pages = {205423},
    volume = {111},
    publisher = {American Physical Society},
    url = {https://journals.aps.org/prb/abstract/10.1103/PhysRevB.111.205423},
    doi = {10.1103/PHYSREVB.111.205423/COMPILED{\_}SUPPLEMENTARY.PDF},
    issn = {24699969}
}

@misc{Tmterud2022TemperatureGraphene,
    title = {{Temperature Dependent Bending Rigidity of Graphene}},
    year = {2022},
    author = {T{\o}mterud, Martin and Hellner, Simen K and Eder, Sabrina D and Forti, Stiven and Manson, Joseph R and Colletti, Camila and Holst, Bodil},
    month = {10},
    publisher = {arXiv},
    url = {http://arxiv.org/abs/2210.17250},
    doi = {10.48550/arXiv.2210.17250}
}

@article{Wang2005SizeNanotubes,
    title = {{Size dependence of the thin-shell model for carbon nanotubes}},
    year = {2005},
    journal = {Physical Review Letters},
    author = {Wang, Lifeng and Zheng, Quanshui and Liu, Jefferson Z. and Jiang, Qing},
    number = {10},
    month = {9},
    pages = {105501},
    volume = {95},
    publisher = {American Physical Society},
    url = {https://journals.aps.org/prl/abstract/10.1103/PhysRevLett.95.105501},
    doi = {10.1103/PHYSREVLETT.95.105501/DELIVERABLE/1D892E12-DB9A-4051-B748-6DA35F65C3EC},
    issn = {00319007}
}

@article{Pine2014VibrationalStrain,
    title = {{Vibrational analysis of thermal oscillations of single-walled carbon nanotubes under axial strain}},
    year = {2014},
    journal = {Physical Review B - Condensed Matter and Materials Physics},
    author = {Pine, Polina and Yaish, Yuval E. and Adler, Joan},
    number = {11},
    month = {3},
    pages = {115405},
    volume = {89},
    publisher = {American Physical Society},
    url = {https://journals.aps.org/prb/abstract/10.1103/PhysRevB.89.115405},
    doi = {10.1103/PHYSREVB.89.115405/DELIVERABLE/BAFFABD4-EF7B-40B2-91AE-8B74D9579519},
    issn = {1550235X}
}

@article{Huang2006ThicknessNanotubes,
    title = {{Thickness of graphene and single-wall carbon nanotubes}},
    year = {2006},
    journal = {Physical Review B - Condensed Matter and Materials Physics},
    author = {Huang, Y. and Wu, J. and Hwang, K. C.},
    number = {24},
    month = {12},
    pages = {245413},
    volume = {74},
    publisher = {American Physical Society},
    url = {https://journals.aps.org/prb/abstract/10.1103/PhysRevB.74.245413},
    doi = {10.1103/PHYSREVB.74.245413/DELIVERABLE/94035831-31D3-411F-BEAC-28BE07DBC62F},
    issn = {10980121}
}

\end{document}